\documentclass[12pt]{book}

\usepackage[dvips]{graphicx,color}
\usepackage{makeidx,tsukuba}
\usepackage{wrapfig}

\makeauthorindex
\makeindex

\begin{document}

\BookTitle{\itshape The 28th International Cosmic Ray Conference}
\CopyRight{\copyright 2003 by Universal Academy Press, Inc.}
\pagenumbering{arabic}

\chapter{
Limitations on Space-based Air Fluorescence Detector Apertures obtained
from IR Cloud Measurements}

\author{%
%
%
John Krizmanic,$^{1,2}$, Pierre Sokolsky$^3$, and Robert Streitmatter$^2$ \\
{\it (1) Universities Space Research Association\\
(2) NASA Goddard Space Flight Center, Greenbelt, Maryland 20771, USA\\
(3) High Energy Astrophysics Institute, University of Utah, Salt Lake City, Utah 84112, USA} \\
}

\section*{Abstract}
The presence of clouds between an airshower and a space-based
detector can dramatically alter the measured signal characteristics
due to absorption and scattering of the photonic signals.  
Furthermore, knowledge of the cloud cover in the observed atmosphere is needed
to determine the instantaneous aperture of such a detector.  Before exploring the
complex nature of cloud-airshower interactions, we examine a simpler issue.
We investigate
the fraction of ultra-high energy cosmic ray events that may be
expected to occur
in volumes of the viewed atmosphere non-obscured by clouds.
To this end, we use space-based IR data in concert
with Monte Carlo simulated $10^{20}$ eV airshowers to determine the
acceptable event fractions. Earth-observing instruments, such as
MODIS, measure detailed cloud
configurations
via a CO$_2$-slicing
technique that can be used to determine cloud-top altitudes over large
areas.
Thus, events can be
accepted if their observed 3-dimensional endpoints occur above low clouds 
as well as from areas of cloud-free atmosphere.  An initial analysis has determined
that by accepting airshowers that occur above low clouds, 
the non-obscured acceptance can be increased by approximately a factor of 3
over that obtained using a cloud-free criterion.

\section{Introduction}

Using the air fluorescence technique, the space-based experiments
EUSO [1] and OWL [4] will image the UV
nitrogen fluorescence signal from extended airshowers with a spatial resolution of
$\sim 1$ km$^2$ (on the ground) while instantaneously monitoring areas approaching
$10^6$ km$^2$.  The effects of clouds
must be well understood as clouds can obscure or
modify the signal strengths and profiles of the airshowers.   The incorporation of
a LIDAR system has been proposed to be used in conjunction with the EUSO and OWL detectors, but the
scanning requirements imposed by the nearly mega-pixel count and the rapidly moving footprint(s)
lead to severe operational requirements.
An alternate approach to determine the cloud properties is the use of meteorological measurements.
The MODIS instruments [3] on the Terra and Aqua satellites provide derived
cloud property measurements from IR measurements
with a spatial resolution of $\sim 1$ km$^2$ on the ground.  Using the technique of CO$_2$ slicing
[7], cloud-top altitudes can be determined and lead to 3-dimensional cloud profiles.
The availability of these and
future meteorological measurements could
be used by space-based air fluorescence experiments to define clear apertures and relax
LIDAR operational constraints.

In a previous paper [5], we used simulated  airshowers superimposed over MODIS
measured scenes and obtained {\it the fraction of tracks that occur in cloud-free portions of the atmosphere}.  This 
analysis
used a cloud-mask product that provides a probabilistic determination of whether a particular 1 km$^2$ pixel
contains a cloud.  Using a conservative selection criteria defined by requiring no cloudy pixels within 3 km of
the observed portion of a simulated airshower, we determined that the mean of the distribution was 6.5\%.
Assuming a 15\% duty cycle imposed by moonless night along with other operational constraints, a 6.5\% clear fraction
leads to a disconcerting 1\% effective duty cycle. 

In this paper, we present the first results on the potential of recovering airshowers with observed endpoints that
occur above low clouds and thus are unobscured.  This is accomplished by 
employing another MODIS data product which yields
cloud-top pressures, i.e.  cloud-top altitudes, but on a courser $5 \times 5$ km$^2$ pixel scale.
The unobscured event fraction
is then determined by combining the cloud-free fraction with the low-cloud fraction. 

\section{Method of Analysis}

A simulated, airshower event sample was generated using the OWL Monte Carlo [2] 
assuming an isotropic flux of $10^{20}$
eV protons.  
The two OWL instruments [6] were configured to be in 1000 km orbits with a 
500 km satellite separation.
The 3-dimensional airshower tracks were reconstructed using the stereo reconstruction
technique with the track endpoints determined by requiring observation by both instruments.  The approximately
1700 tracks were then superimposed upon 78 different MODIS scenes defined by the MODIS Collection4
cloud-top product.  Each pixel was coded with either being clear, having a
cloud top with an altitude of 3 km or less, or a cloud top with an altitude higher than 3 km.  The spatial location
of the xy-projected track lengths were then compared to the MODIS cloud pixel information to determine whether
the tracks occurred in a clear, low-cloud, or high-cloud portion of the MODIS scene.  The extended nature of
the airshowers will lead to a reduced acceptance as compared to that expected by determining the cloud-free
or low-cloud pixel fractions.

\begin{figure}[h]
\vfill \begin{minipage}[t]{7cm}
\begin{center}
\includegraphics[height=17.5pc]{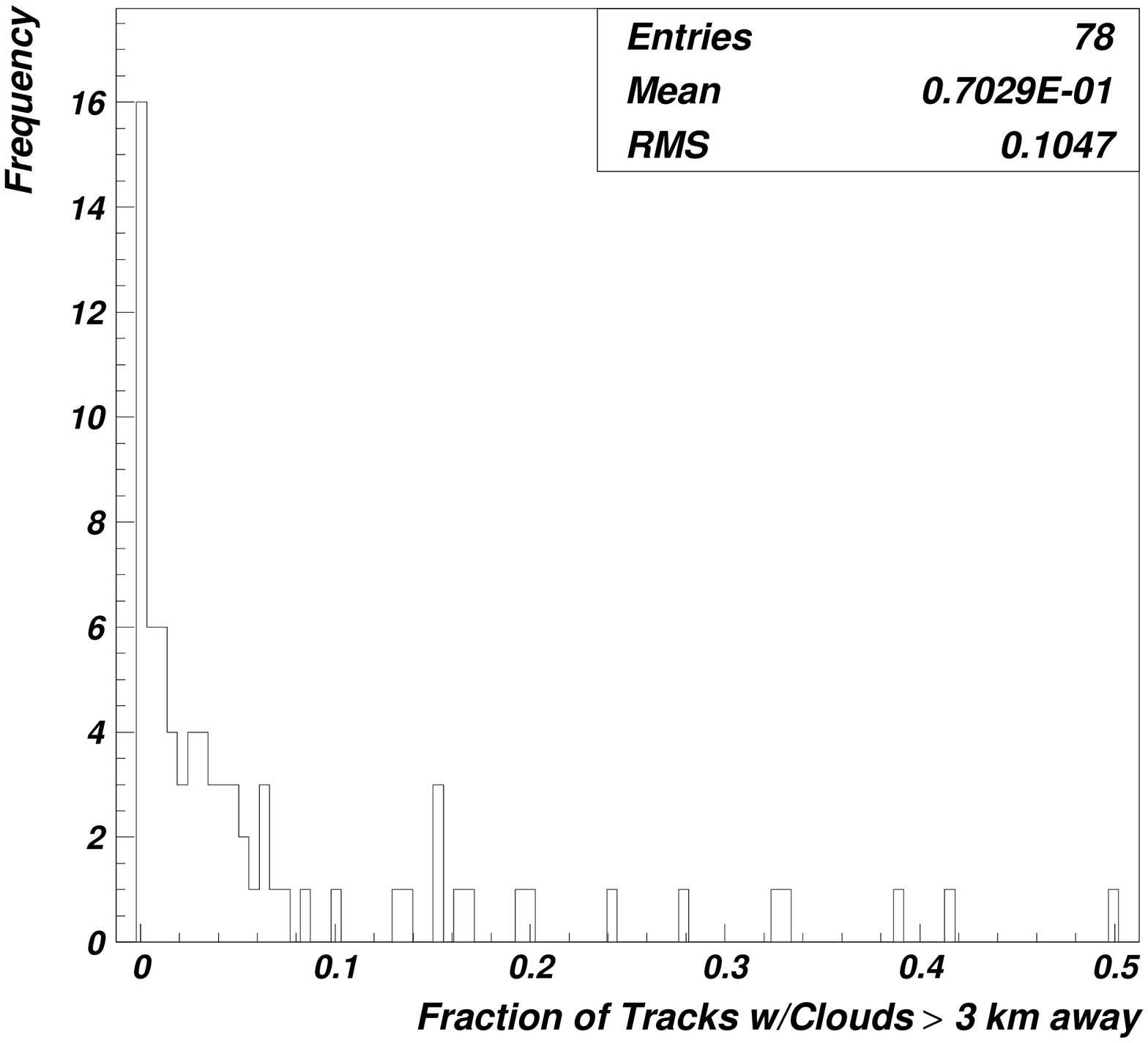}
\end{center}
\caption{Distribution of fractional clear aperture obtained from the 1 km$^2$ MODIS cloud-mask data.}
\end{minipage} \hfill
\begin{minipage}[t]{7cm}
\begin{center}
\includegraphics[height=17.5pc]{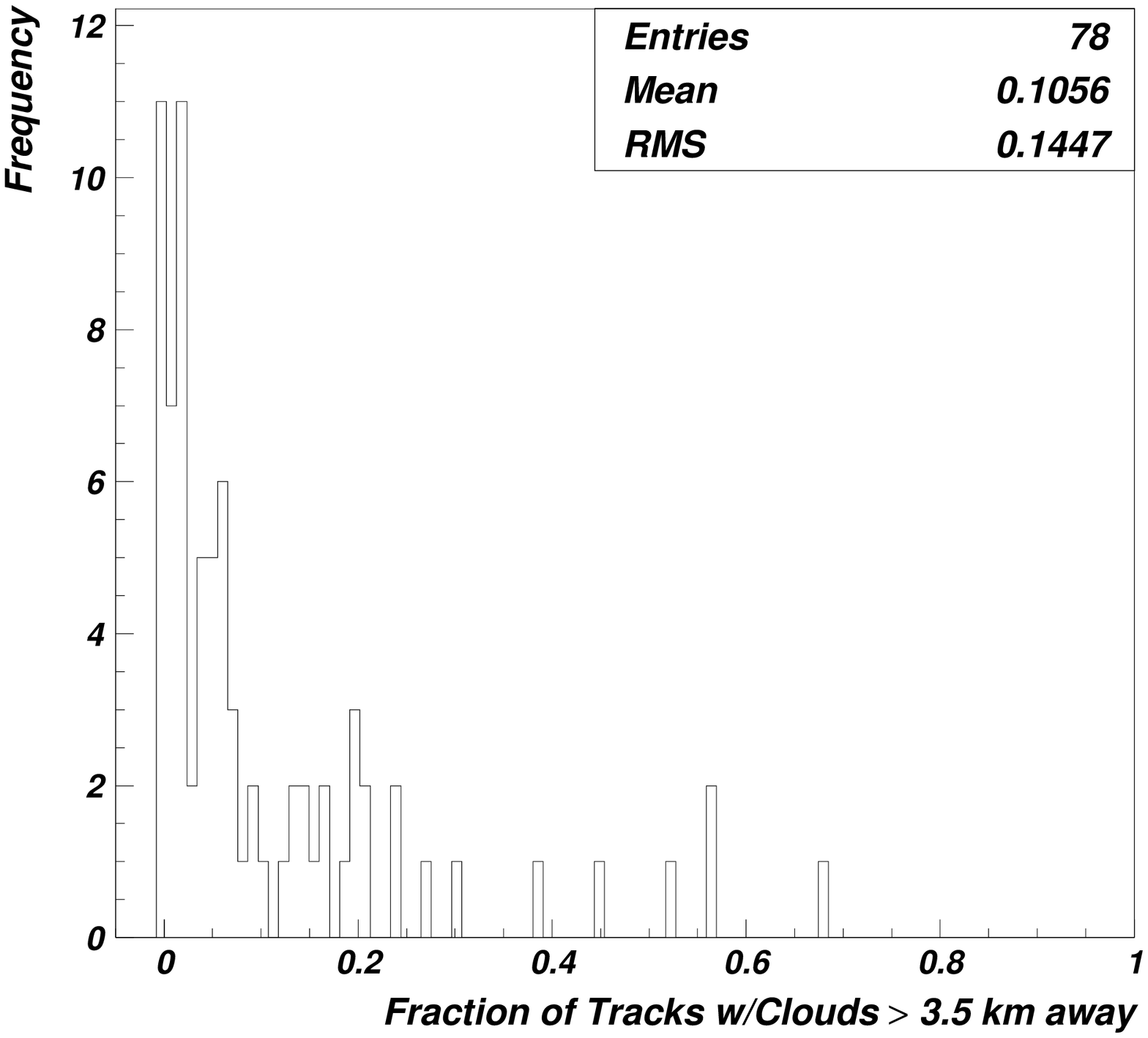}
\end{center}
\caption{Distribution of fractional clear aperture obtained from the $5 \times 5$ km$^2$ MODIS cloud-top
pressure data.}
\end{minipage}
\vfill
\end{figure}

\begin{wrapfigure}{r}{8.1cm}
\vspace{-1.9cm}
\includegraphics[height=20.pc]{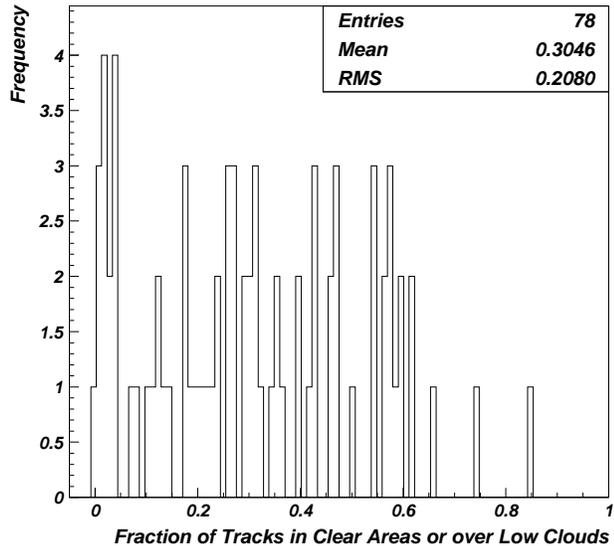}
\caption{Distribution of fractional clear and low-cloud aperture obtained from the $5 \times 5$ km$^2$ MODIS cloud-top
pressure data.}
\end{wrapfigure}
\section{Results}

Figure 1 shows the distribution of the cloud-free track fraction using the 1 km$^2$ cloud-mask product for the 78 MODIS
scenes located near the equator used in this study.  The analysis follows that employed in our previous paper [5]: a track is considered
in a cloud-free portion of the atmosphere if no cloudy pixels are within 3 km of the observed track.  The results presented
in Figure 1 employed a newer version, Collection4, of the cloud-mask determining algorithm.  The mean of 7\% matches
well to the 6.5\% we had obtained for a slightly larger MODIS data set by using a previous version of the
MODIS cloud-mask algorithm.  The 78 scenes used in this study are a subset of the 85 scenes used in the
original study.  Figure 2 shows the distribution of cloud-free track fractions for the 78 MODIS scenes using the
MODIS cloud-top data product.  
The cloud-free fractions were determined by requiring no cloudy pixels to be
within 3.5 km of the track.  The slightly larger value of 3.5 km is imposed by the courser $5 \times 5$ km$^2$ pixel size 
used in the cloud-top product.  Note that the cloud-free fraction of 10.7\% obtained by this analysis is in relatively
good agreement with the 7\% obtained from the 1 km$^2$ cloud-mask analysis.

Figure 3 shows the distribution of the fraction of tracks that occur over either cloud-free areas or have a portion of the xy-projected track over clouds with cloud-top heights of
less than 3 km.  
A track over low clouds was accepted if it was at least 3.5 km from a pixel with a $> 3$ km altitude cloud and has an observed track
endpoint at an altitude 4 km or higher.  
The simulated event sample has 53\% of the tracks with observed endpoints above
4 km in altitude.  The distribution in Figure 3 exhibits a mean of over 30\% and is flatter than that for
the cloud-free fraction (Figure 2).  The 30\% value is a factor of 3 larger than
that for the cloud-free fraction.  If the track endpoint requirement is increased to 5 km for low-cloud acceptance, the mean
of the combined cloud-free and low-cloud distribution is reduced to 26\%.

\section{Discussion}

The ability to determine the altitudes of airshowers and cloud tops allows for a significant increase in the acceptance of
events by including tracks whose observable endpoints occur above low clouds.   The potential for the
increase is echoed in the 16\% cloud-free versus 53\% cloud-free+low-cloud pixel fractions obtained from
the MODIS cloud-top data used in this study.

We would like to thank Bill Ridgway for invaluable discussions and for the generation of the MODIS
data files. 

\section{References}

\re
1.\ http://www.euso-mission.org/ \
\re
2.\ Krizmanic, J. for the OWL Collaboration\ 2001, Proc. 27th ICRC, Vol. 2, 861
\re
3.\ http://modis-atmos.gsfc.nasa.gov/\
\re
4.\ http://owl.gsfc.nasa.gov/\
\re
5.\ Sokolsky P., Krizmanic J.\ 2003, astro-ph/0302501, submitted to Astroparticle Physics
\re
6.\ Streitmatter R. et al.\ 2002, http://owl.gsfc.nasa.gov/WP10c.PDF
\re
7.\ Stumpf R., Pennock J.\ 1989, Jour. Geophys. Res. 94(C10), 14,363

\endofpaper
\end{document}